# TM-polarized Surface Plasmon Polaritons in Nonlinear Multi-layer Graphene-Based Waveguides: An Analytical Study

Mohammad Bagher Heydari [1], Mohammad Hashem Vadjed Samiei [1,*]

[1,*] School of Electrical Engineering, Iran University of Science and Technology (IUST), Tehran, Iran

[*]Corresponding author: mh_samiei@iust.ac.ir

**Abstract:** This paper presents an analytical study of TM-polarized surface plasmon polaritons (SPPs) in nonlinear multi-layer structures containing graphene sheets. In the general structure, each graphene sheet has been sandwiched between two different nonlinear magnetic materials. To show the richness of the proposed general waveguide, two novel nonlinear structures have been introduced and investigated as special cases of the general structure. It will be shown that the propagation features of these structures can be tuned by changing the chemical potential of the graphene and the incident mode power. A large value of the effective index, i.e. $n_{eff} = 240$ for the chemical potential of $\mu_c = 0.2\ ev$ and the incident power of $\alpha|H_{y,0}|^2 = 3$ is obtained for the second structure at the frequency of 40 THz. The analytical results confirm that the integration of nonlinear magnetic materials with graphene sheets can control and enhance the propagating features and the self-focused of the field in the nonlinear layer. This integration gives more degrees of freedom to the designer to propose new THz components such as lasers and switches in the THz region.

## 1. Introduction

Graphene plasmonics, one of the interesting research areas among scientists, has been emerged and developed based on the propagating SPPs on the graphene sheets [1]. The ability of high mode confinement in the THz frequencies and tunability of SPPs make the graphene a promising two-dimensional material to be utilized in a large number of THz devices such as couplers [2], sensors [3,4], filters [5,6], and waveguides [7-14]. It should be noted that noble metals also support SPPs at the near-infrared and visible frequencies [15-17]. However, some significant features of graphene-based waveguides, such as extreme confinement, tunable conductivity, and low losses in THz and mid-infrared frequencies differ from any metal-air interface waveguides [18,19].

One way to effectively enhance the performance of graphene-based devices is by integrating it with advanced materials such as anisotropic and nonlinear materials. The hybridization of graphene with other materials gives more degrees of freedom to the designer to control and enhance the propagating properties. One of the interesting materials to be integrated with graphene is nonlinear material, which exhibits some fascinating properties such as strong field confinement and optical bi-stability [20-22]. Two types of nonlinear materials have attracted the attention of the scientists to integrate them with graphene-based devices: 1- Kerr-type materials, 2- Nonlinear magnetic materials. For instance, a Kerr-type medium has been integrated with graphene layers to improve the propagation properties in [23,24]. In [25], a Kerr-type dielectric slab bounded by two graphene sheets has been studied numerically to adjust the plasmonic features by changing the nonlinear parameter and the width of the structure. The integration of a nonlinear magnetic medium with a graphene sheet has been investigated in [26], where a graphene layer has been sandwiched between a dielectric and a nonlinear magnetic material.

In this paper, an analytical model is proposed for nonlinear multilayer waveguides incorporating graphene sheets. In the general structure, each graphene sheet has been sandwiched between two different nonlinear magnetic materials. Each nonlinear magnetic material in the N-th layer has the permittivity and the permeability of $\varepsilon_N^L, \mu_N^{NL}$, respectively. It should be noted that nonlinear magnetic materials such as $FeF_2$ and $MnF_2$ exhibit interesting properties that can enhance the performance of THz devices [27-31]. To the authors' knowledge, no published work has been reported on the analytical study of nonlinear multilayer graphene-based structures with nonlinear magnetic materials. The presented study can give a new platform for designing innovative plasmonic devices in the THz frequencies.

The paper is organized as follows. Section 2 introduces the general structure and proposes its analytical model. A nonlinear differential equation will be solved in this section to obtain the components of TM waves. In section 3, two novel nonlinear waveguides, as special cases of the general structure, are introduced and investigated. To the authors' knowledge, these proposed waveguides have not been studied in any published article. The analytical results will show that the hybridization of the graphene sheet with nonlinear magnetic materials can enhance propagation and localization characteristics. Finally, section 4 concludes the article.

## 2. The Proposed Structure and its Analytical Model

Fig. 1 illustrates the schematic of the general structure, where each graphene sheet has been sandwiched between two different nonlinear magnetic materials. Each nonlinear magnetic material in the N-th layer has the permittivity and the permeability of $\varepsilon_N^L, \mu_N^{NL}$, respectively, where the permeability of nonlinear magnetic material is expressed as [30]:

$$\mu_N^{NL}(\omega) = \mu_N^L(\omega) + \alpha_N |\boldsymbol{H}|^2 \tag{1}$$

Where

$$\mu_N^L(\omega) = 1 + \frac{2\omega_{m,N}\omega_{a,N}}{\omega_{c,N}^2 - \omega^2} \tag{2}$$

$$\omega_{m,N} = \mu_0 \gamma_N M_{s,N} \tag{3}$$

$$\omega_{a,N} = \mu_0 \gamma_N H_{A,N} \tag{4}$$

$$\omega_{e,N} = \mu_0 \gamma_N H_{E,N} \tag{5}$$

$$\omega_{c,N}^2 = \omega_{a,N}^2 + 2\omega_{a,N}\omega_{e,N} \tag{6}$$

In the above equations, $\gamma_N$ is the gyromagnetic ratio, $\omega_{c,N}$ is the resonance frequency, $M_{S,N}$ is the saturation magnetization field, $H_{A,N}$ is the anisotropy field and $H_{E,N}$ is the exchange field of the nonlinear magnetic material in the N-th layer. It should be noted that *NL, L,* and *N* denote "Non-linear", "Linear" and N-th layer of the structure. Furthermore, the conductivity of the graphene in the *N*-th layer has the following well-known relation [32]:

$$\sigma_N(\omega, \mu_{c,N}, \Gamma_N, T) = \frac{-je^2}{4\pi\hbar} Ln\left[\frac{2|\mu_{c,N}| - (\omega - j2\Gamma_N)\hbar}{2|\mu_{c,N}| + (\omega - j2\Gamma_N)\hbar}\right] + \frac{-je^2 K_B T}{\pi\hbar^2(\omega - j2\Gamma_N)}\left[\frac{\mu_{c,N}}{K_B T} + 2Ln\left(1 + e^{-\mu_{c,N}/K_B T}\right)\right] \tag{7}$$

Where $\hbar$ is the reduced Planck's constant, $K_B$ is Boltzmann's constant, $\omega$ is radian frequency, $e$ is the electron charge, $\Gamma_N$ is the phenomenological electron scattering rate for that layer ($\Gamma_N = 1/\tau_N$, where $\tau_N$ is the relaxation time), *T* is the temperature, and $\mu_{c,N}$ is the chemical potential for the N-th layer which can be altered by chemical doping or electrostatic bias [32].

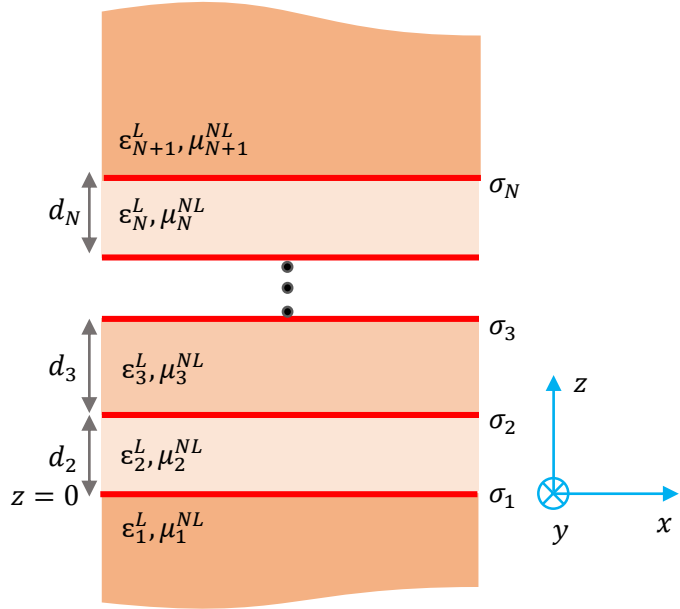

**Fig. 1.** The schematic of the general structure. In this waveguide, each graphene sheet has been sandwiched between two different nonlinear magnetic materials.

Consider Maxwell's equations inside the nonlinear magnetic material of the N-th layer in the frequency domain (suppose $e^{-i\omega t}$):

$$\nabla \times \boldsymbol{E}_N = j\omega\mu_0\mu_N^{NL}\boldsymbol{H}_N \tag{8}$$

$$\nabla \times \boldsymbol{H}_N = -j\omega\varepsilon_0\varepsilon_N^L\boldsymbol{E}_N \tag{9}$$

Let us suppose that TM-polarized plasmonic waves ($H_y(z), E_x(z), E_z(z)$) propagating in the x-direction ($e^{i\beta x}$) inside the nonlinear magnetic material. By substituting the components of TM waves, i.e. $H_y(z), E_x(z), E_z(z)$, in (8)-(9), we obtain the following equation for the N-th layer of nonlinear magnetic material:

$$\frac{d^2 H_{y,N}}{dz^2} - \gamma_N^2 H_{y,N} + k_0^2 \varepsilon_N^L \alpha_N H_{y,N}^3 = 0 \tag{10}$$

Where $k_0$ is the free-space wave-number and,

$$\gamma_N^2 = \beta^2 - k_0^2 \varepsilon_N^L \mu_N^L \tag{11}$$

The transverse component of electric fields can be obtained as:

$$E_{x,N} = \frac{1}{j\omega\varepsilon_0\varepsilon_N^L}\frac{\partial H_{y,N}(z)}{\partial z} \tag{12}$$

$$E_{z,N} = \frac{-\beta}{\omega\varepsilon_0\varepsilon_N^L} H_{y,N}(z) \tag{13}$$

Now, the nonlinear equation of (10) should be solved to obtain the dispersion relation and field distributions of all nonlinear layers. The first integration of (10) yields to

$$\left[\frac{d H_{y,N}}{dz}\right]^2 - \gamma_N^2 H_{y,N}^2 + \frac{1}{2} k_0^2 \varepsilon_N^L \alpha_N H_{y,N}^4 = C_N \tag{14}$$

Where $C_N$ is a constant of integration and,

$$C_N = \left[\frac{d H_{y,N}}{dz}\right]^2 \bigg|_{z = z_N = \sum_{i=2}^{N-1} d_i} - \gamma_N^2 H_{y,N}^2(z_N) + \frac{1}{2} k_0^2 \varepsilon_N^L \alpha_N H_{y,N}^4(z_N) \tag{15}$$

Let us assume that $C_N \geq 0$ (for the case of $C_N \leq 0$, the expressions are similar). Now, equation (14) is expressed as follows:

$$\frac{d H_{y,N}}{dz} = \pm\sqrt{C_N + \gamma_N^2 H_{y,N}^2 - \frac{1}{2} k_0^2 \varepsilon_N^L \alpha_N H_{y,N}^4} \tag{16}$$

By integrating (16) and doing some rigorous mathematical procedures, the solution of (16) is obtained in general form for the N-th layer:

$$H_{y,N}(z) = A_N \, cn\left(p_N\left[(z - z_N) + z_{oc,N}\right], m_N\right) \tag{17}$$

Where the following parameters have been defined in (17):

$$p_N = \sqrt[4]{\gamma_N^4 + 2C_N k_0^2 \varepsilon_N^L \alpha_N} \tag{18}$$

$$m_N = \frac{p_N^2 + \gamma_N^2}{2 p_N^2} \tag{19}$$

$$A_N = \frac{1}{k_0}\sqrt{\frac{p_N^2 + \gamma_N^2}{\alpha_N}} \tag{20}$$

In (17), "$cn$" is Jacobi elliptic function and $m_N$ is Jacobi modulus [33]. The general solution of relation (17) has some special cases. One of the important specific cases occurs for $C_N = 0, m_N = 1$, which results in

$$H_{y,N}(z) = \frac{\gamma_N}{k_0}\sqrt{\frac{2}{\alpha_N}} \, sech\left(\gamma_N\left[(z - z_N) + z_{oc,N}\right]\right) \tag{21}$$

In general form, the boundary conditions for a graphene sheet sandwiched between two nonlinear magnetic materials are written as:

$$E_{x,N+1} = E_{x,N} \quad , \quad H_{y,N+1} - H_{y,N} = -\sigma_N E_{x,N} \tag{22}$$

By utilizing the relations of (12)-(13) and (17)-(22), the dispersion relation is obtained. Now, achieving other propagating properties such as the effective index is straightforward.

## 3. Special Cases of the Proposed Structure: Results and Discussions

To show the richness of the proposed general structure, two new graphene-based waveguides are considered in this section. The first structure is a nonlinear slab waveguide containing a graphene sheet. In this structure, a graphene layer has been deposited on $FeF_2$-$SiO_2$-Si layers. The second one is a symmetric nonlinear waveguide, where two nonlinear layers at both sides of the whole waveguide have been utilized as cladding layers. It will be shown that the usage of two graphene sheets together with two nonlinear materials enhances the performance of the structure.

It should be mentioned that the thickness of the graphene layer can change and enhance the propagation features of the waveguide such as propagation length because this parameter appears in the characteristic equation of the waveguide by applying boundary conditions (see relation (22)). Furthermore, the effective plasma frequency (defined by $\omega_p^{eff} = e\sqrt{\frac{n_c}{m_c \varepsilon_0 \Delta}}$ where $n_c, m_c, \Delta$ are the carrier concentration, the effective mass of fermions, and the thickness of guiding graphene layer) depends on the thickness of graphene. In all analytical results, graphene layers are at the temperature of $T = 300\ K$, have a thickness of $\Delta = 0.33\ nm$, and relaxation time of $\tau = 100\ fs$. The $SiO_2$ and Si layers have the permittivities of 2.09 and 11.9, respectively. Here, we suppose that the nonlinear magnetic material is $FeF_2$ with the following parameters: $\gamma = 1.76 \times 10^{11}\ (Ts)^{-2}$, $M_S = 4.46 \times 10^4\ \frac{A}{m}$, $H_A = 1.59 \times 10^4\ \frac{A}{m}$, $H_E = 4.3 \times 10^4\ \frac{A}{m}$, $\alpha = 8.869 \times 10^{-8}\ m^2 A^{-2}$, $\varepsilon_{FeF_2} = 4$. The geometrical parameters in all structures have been considered $t = 0.2\ \mu m, d = 0.5\ \mu m, s = 0.6\ \mu m$.

*3.1 The First Structure: A Nonlinear Slab Waveguide Containing a Graphene sheet*

Fig. 2 illustrates the schematic of the first structure, where a graphene layer has been deposited on $FeF_2$-$SiO_2$-Si layers. In this structure, the incident mode power (or power of the incident beam) is supposed to be $\alpha |H_{y,0}|^2 = 3$ unless otherwise stated.

The propagation properties of the first structure as a function of the frequency have been shown in Fig. 3. In Fig. 3 (a), the effective index has been defined as $n_{eff} = Re[\beta]/k_0$. The propagation length ($L_{Prop}$) is normalized with plasmon wavelength ($L_{Normalized,Prop} = L_{Prop}/\lambda_{sp}$) in Fig. 3 (b). One can observe that the normalized propagation length increases as the frequency increases. It can be seen from Fig. 3 (a) that at a specific frequency (for instance, consider $f = 40\ THz$), the effective index reduces as the chemical potential increases.

To study the effect of nonlinear magnetic coefficient and also the chemical potential of the graphene on the effective index, we have depicted Fig. 4. As seen in Fig. 4 (a), the effective index increases as the incident power increases. However, the slope of variations is small. Furthermore, higher values of the chemical potential obtain lower values of the effective index, which means that for a specific value of incident power, the effective index decreases as the chemical potential increases. The main reason for this matter is the reduction of field concentration on the graphene sheet for higher values of chemical potential.

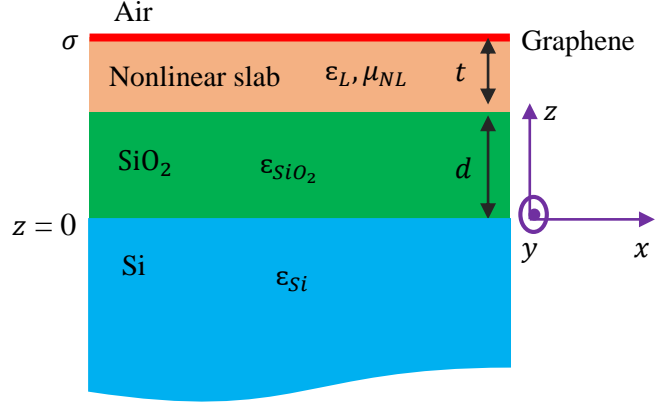

Fig. 2. The schematic of the first structure: a graphene sheet has been located on FeF$_2$-SiO$_2$-Si layers.

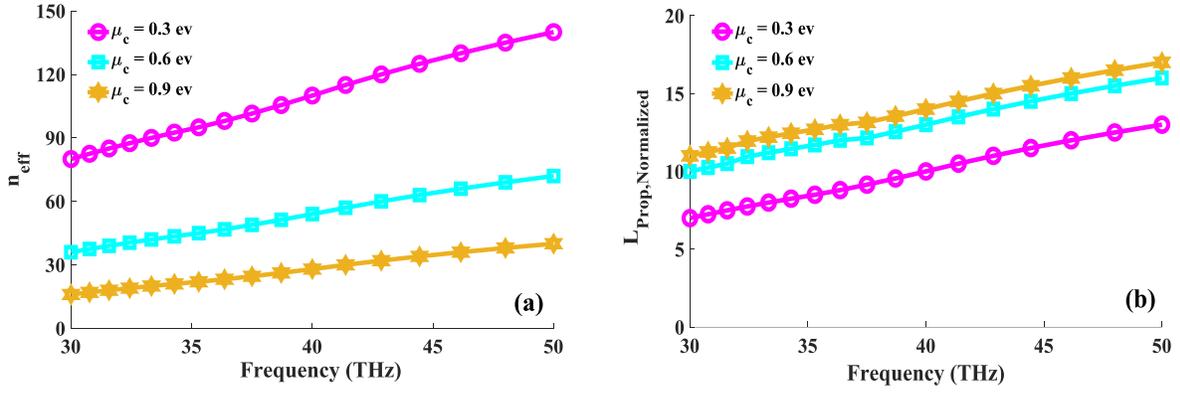

Fig. 3. The propagating properties of the first structure as a function of the frequency for various values of the chemical potential. The incident mode power is $\alpha|H_{y,0}|^2 = 3$.

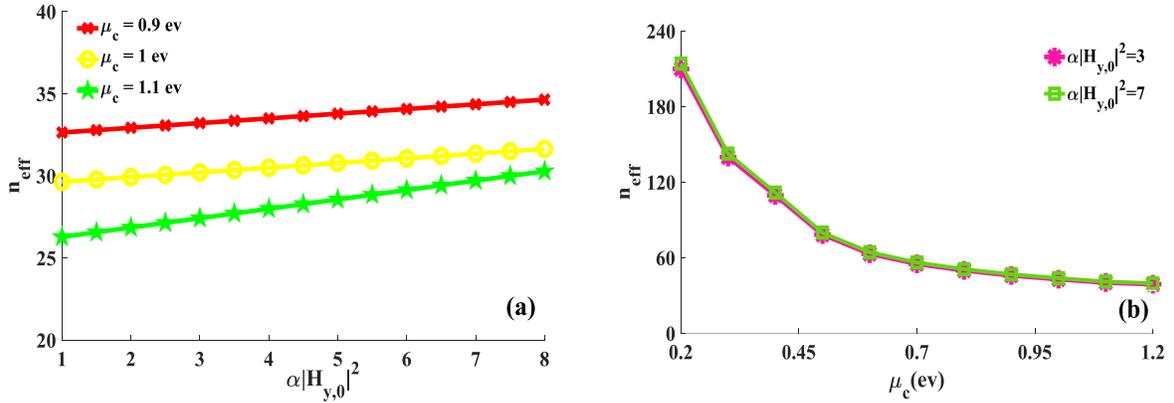

Fig. 4. The analytical results of the effective index as a function of: (a) the incident mode power, for various values of the chemical potential, (b) the chemical potential for various values of incident mode power. The operation frequency is 40 THz.

Fig. 4 (b) illustrates the effective index of the first structure as a function of chemical potential for various values of incident power. It is evident that the effective index decreases with the increment of chemical potential. To explain this matter, it is worthwhile to be mentioned that the graphene layer acts as a conductor when its conductivity increases with the increment of the chemical potential. Therefore, the electromagnetic fields concentrate on SiO$_2$-Si layers, and thus the effective index decreases. Another important point in Fig. 4 (b) is the independence of this diagram on the incident power. It is observed from Fig. 4 (b) that the effective index varies slightly for various values of incident power. This matter confirms the analytical results of Fig. 4 (a), where the effective index changes slightly with the increment of incident mode power.

As a final point for this sub-section, the distribution of normalized H$_y$ has been depicted in Fig. 5 at the frequency of 40 THz for the chemical potential of $\mu_c = 0.3\ ev$ and the incident mode power of $\alpha|H_{y,0}|^2 = 3$. It can be observed that the most energy is concentrated on the FeF$_2$-Graphene interface and self-peak in nonlinear magnetic slab occurs at $z \approx 0.55\ \mu m$. The first structure showed high values of the effective index and strong nonlinear effects. Also, the main advantage of the studied case was its ability to control the propagation features by changing chemical potential. In what follows, we will design and study a hybrid symmetric graphene-based waveguide, which can support highly-concentrated SPPs and thus will give a more tunable effective index.

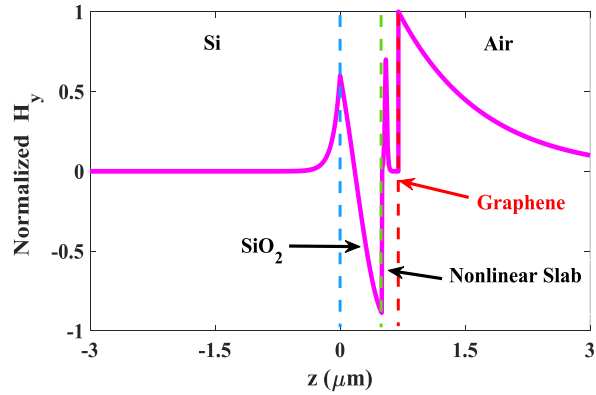

Fig. 5. The normalized H$_y$ distribution at the frequency of 40 THz. The chemical potential and the power of the incident beam are supposed to be $\mu_c = 0.3\ ev, \alpha|H_{y,0}|^2 = 3$, respectively.

*3.2 The Second Structure: A Symmetric Graphene-based Waveguide with Two Nonlinear Magnetic Layers*

As a second case, we study a hybrid symmetric graphene-based waveguide with two nonlinear layers. The schematic of this structure has been represented in Fig. 6, where two nonlinear magnetic layers have been utilized as cladding layers. Similar to the first structure, the nonlinear magnetic medium is chosen FeF$_2$ and its parameters have been given previously. Both graphene layers have similar parameters as mentioned before.

Fig. 7 demonstrates the propagation properties of the second structure as a function of the frequency for various values of the chemical potential. As seen in this figure, the effective index and the normalized propagation length increases as the frequency increases. For higher values of the chemical potential, the effective index reduces but the normalized propagation length increases. Therefore, there is a trade-off for choosing a suitable chemical potential to achieve the desired values of the effective index and the propagation length. Compared to the plasmonic features of the first structure, this structure exhibits higher values of the effective index and the propagation length, which originates from the integration of two graphene layers with two nonlinear magnetic layers.

One of the important parameters for designing the structure is the incident mode power, which can control the nonlinear effects of the structure. As seen in Fig. 8 (a), the effective index increases slightly as the incident beam increases. However, the slope of the variations increases for higher values of the chemical potential. For instance, the slope of the variations for $\mu_c = 1.1\ ev$ increases compared to $\mu_c = 0.9\ ev, 1\ ev$. Therefore, the incident power has a great influence on the propagation parameters for higher values of chemical potential. It is evident from Fig. 8(b) that the normalized propagation length has a constant value for various values of the chemical potential for the range $\alpha|H_{y,0}|^2 < 4$. However, as the incident mode power increases, the value of chemical potential influences the normalized propagation length. As a result, for achieving a large value of the propagation length, one should design the second structure with a higher value of chemical potential and incident power.

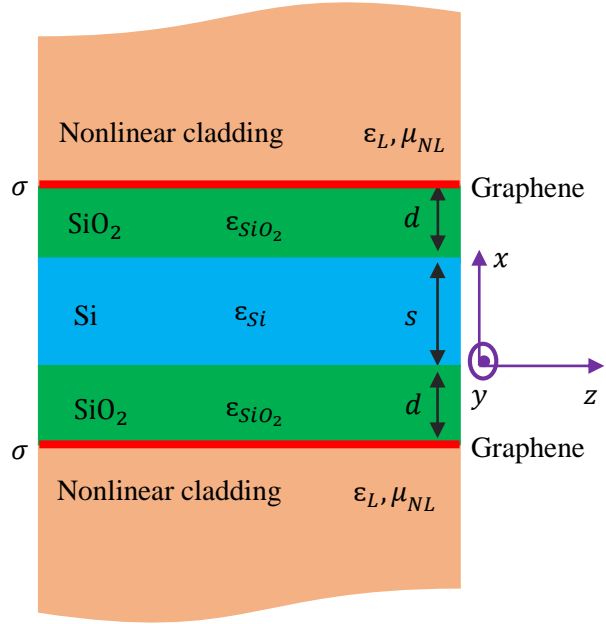

Fig. 6. The schematic of the second structure: a symmetric nonlinear waveguide, where two nonlinear layers at both sides of the whole waveguide have been utilized as cladding layers.

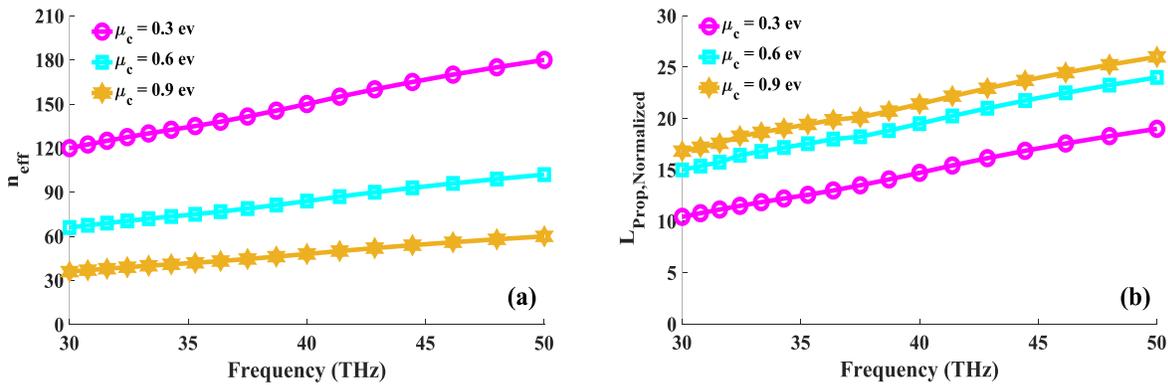

Fig. 7. The propagating properties of the second structure as a function of the frequency for various values of the chemical potential. The incident mode power is $\alpha|H_{y,0}|^2 = 3$.

To see the effect of the chemical potential on the propagation features, we have depicted Fig. 9. One can observe from Fig. 9 (a) that the effective index decreases as the chemical potential increases. Compared to the first structure, higher values of the effective index is achievable for this waveguide. For instance, at the chemical potential of $\mu_c =$ $0.2\ ev$, the effective index has the value of $n_{eff} = 200$ for the first structure but it reaches to $n_{eff} = 240$ for the second one. As seen in Fig. 9 (b), the propagation length increases with the increment of the chemical potential. It happens because the electromagnetic energy spreads into the $SiO_2$-Si layers for higher values of the chemical potential. Furthermore, as seen in Fig. 9 (b), the incident mode power has no effect on normalized propagation length for the range of $\mu_c < 0.7\ ev$. Therefore, for graphene layers with higher values of chemical potential, the incident beam should be adjusted for obtaining the desired value of the propagation length. As a final point, it should be mentioned that the integration of two graphene sheets and nonlinear magnetic layers gives a strong nonlinear effect and high values of propagation features. The proposed second waveguide can be utilized for a new generation of THz devices such as lasers and switches.

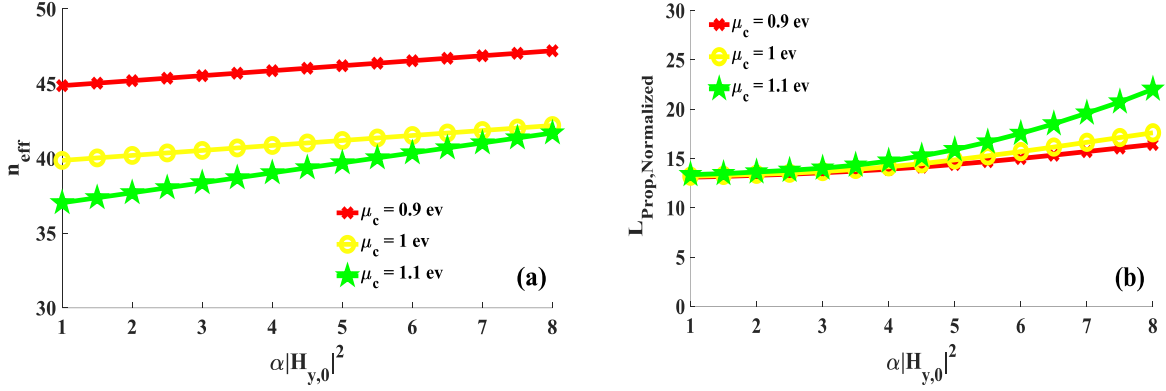

Fig. 8. The analytical results of: (a) the effective index, (b) the normalized propagation length, for the second structure as a function of the incident mode power for various values of the chemical potential. The operation frequency is 40 THz.

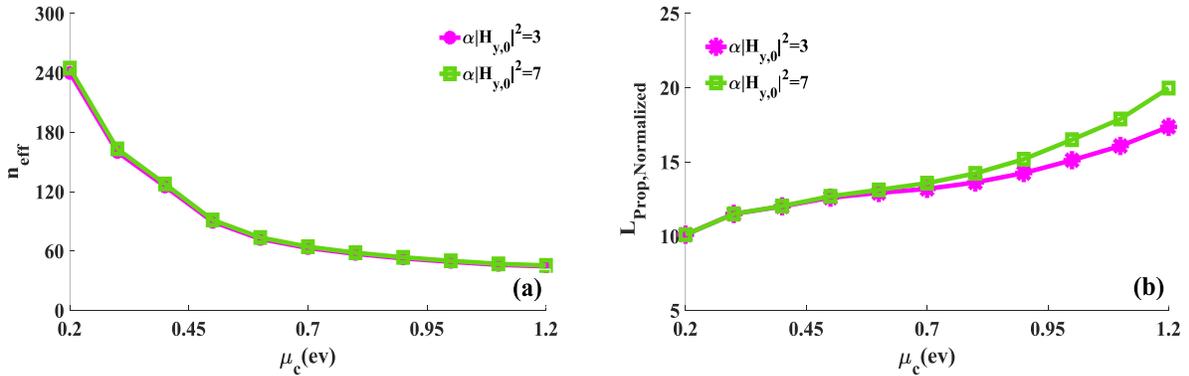

Fig. 9. The analytical results of: (a) the effective index, (b) the normalized propagation length, for the second structure as a function of the chemical potential for various values of the incident mode power. The operation frequency is 40 THz.

## 4. Conclusion

In this article, the analytical study of TM-polarized SPPs in nonlinear multilayer graphene-based waveguides is presented. Two novel structures, as special cases of the general waveguide, are introduced and studied. The first one is a graphene-based nonlinear structure, where a graphene sheet is located on $FeF_2$-$SiO_2$-Si layers. A large value of the effective index, i.e. $n_{eff} = 200$ for the chemical potential of $\mu_c = 0.2\ ev$ and the incident power of $\alpha|H_{y,0}|^2 =$ 3 is obtained at the frequency of 40 THz. The second structure is a hybrid symmetric graphene-based waveguide with two nonlinear magnetic layers. Strong nonlinear effects and high values of the propagation features are achieved for this waveguide, which shows that the usage of two nonlinear magnetic layers and two graphene sheets can enhance the performance of the structure. Furthermore, it is shown in this article that the plasmonic properties of the mentioned waveguides are tunable via the chemical potential of the graphene. The presented analytical study can be utilized for designing new plasmonic devices such as lasers in the THz frequencies.

## Declarations

**Ethics Approval:** Not Applicable.

**Consent to Participate:** Not Applicable.

**Consent to Publish:** Not Applicable.

**Authors Contributions:** M.B.Heydari performed the analytical modeling, conducted the numerical simulations by MATLAB and wrote the manuscript. M.H. Vadjed Samiei supervised the project and reviewed the manuscript.

**Funding:** The authors received no specific funding for this work.

**Competing Interests:** The authors declare no competing interests.

**Availability of Data and Materials:** Not Applicable.